\documentclass[usegraphicx]{mn2e}
\usepackage{times}
\newlength{\colwidth}
\begin{document}
\title{The luminosity dependence of the Type 1 AGN fraction}
\author[Chris Simpson]{Chris Simpson\thanks{Email:
Chris.Simpson@durham.ac.uk} \\
Department of Physics, University of Durham, South Road, Durham DH1 3LE}
\maketitle

\begin{abstract}
Using a complete, magnitude-limited sample of active galaxies from the
Sloan Digital Sky Survey (SDSS) we show that the fraction of broad-line
(Type~1) active galactic nuclei increases with luminosity of the
isotropically-emitted [O{\sc~iii}] narrow emission line. Our results are
quantitatively in agreement with, and far less uncertain than, similar
trends found from studies of X-ray and radio-selected active
galaxies. While the correlation between broad-line fraction and luminosity
is qualitatively consistent with the receding torus model, its slope is
shallower and we therefore propose a modification to this model where the
height of the torus increases slowly with AGN luminosity. We demonstrate
that the faint-end slope of the AGN luminosity function steepens
significantly when a correction for `missing' Type~2 objects is made and
that this can substantially affect the overall AGN luminosity density
extrapolated from samples of more luminous objects.
\end{abstract}

\section{Introduction}

Active galactic nuclei (AGNs) were once thought to be a rare phenomenon
whose major cosmological impact was in their ability to be seen out to
great distances. However, the discovery of supermassive black holes in the
centres of nearby galaxies has led to the view that all galaxies went
through an active phase. Furthermore, calculations have shown that the
energy produced during this phase is sufficient to drive material out of
the host galaxy (Silk \& Rees 1998; Fabian 1999). Feedback such as this can
explain the tightness of the correlation between the masses of the black
hole and host galaxy bulge (e.g., McLure \& Dunlop 2002), and it has been
suggested that such a process may perform a key role in shaping the galaxy
luminosity function (Benson et al.\ 2003).

Semi-analytic models of galaxy formation (e.g., GALFORM; Cole et al.\ 2000)
are now being adapted to investigate the role of AGN-driven feedback, and
determine whether a self-consistent picture of both AGN and galaxy
formation and evolution can be developed. It is to be hoped that progress
in this field may provide important insights into some of the major
outstanding issues in AGN research and, in particular, the question of AGN
triggering. Early attempts at this (e.g., Kauffmann \& Haehnelt 2000) have
appeared encouraging in their ability to fit both the galaxy and QSO
luminosity functions but, by failing to make the distinction between the
\textit{QSO\/} and \textit{AGN\/} luminosity functions, they fail to
account for the possibly significant population of ``Type 2 QSOs'' which
have been postulated to explain the shape of the Cosmic X-ray Background
(CXB; e.g., Comastri et al.\ 1995).

The cosmological importance of these objects is only exceeded by their
elusiveness, since their optical continua closely resemble those of normal
galaxies and therefore they cannot be identified by the colour-selection
techniques used for QSOs (e.g., Schmidt \& Green 1983). Radio selection has
allowed the radio-loud population of Type 2 QSOs (i.e., radio galaxies) to
be identified, but radio sources comprise only a small fraction of the
total AGN population. Hard X-ray selection has failed to locate the
luminous narrow-line AGN in the numbers predicted by model fits to the CXB
but current X-ray telescopes are unable to find distant AGN with
Compton-thick absorption, and such objects are known to exist in abundance
locally (e.g., Risaliti, Maiolino \& Salvati 1999). Large spectroscopic
surveys such as the Sloan Digital Sky Survey (SDSS; York et al.\ 2000) make
it possible to identify AGN in an unbiased manner and therefore determine
for the first time the fraction of Type 2 AGN.

In the simple unified model for AGN, Type~1 and Type~2 objects (i.e., those
which display broad permitted lines, and those which show narrow lines
only, respectively) differ only in terms of the angle which the observer's
line of sight makes with the axis of a dusty torus (see Antonucci 1993 for
more details). If this angle is larger than some critical angle,
$\theta_{\rm c}$, the line of sight to the central regions (which includes
the broad line region) is blocked by the torus and a Type~2 AGN is
seen. QSOs are therefore only a subset of the AGN population and no
cosmological distinction should be made. If $\theta_{\rm c}$ has no
dependence on luminosity, then the QSO and AGN luminosity functions differ
only in terms of their space density normalization.

However, the almost identical SEDs of QSOs just longward of 1\,$\umu$m
(e.g., Kobayashi et al.\ 1993; Elvis et al.\ 1994) imply that the location
of the inner wall of the torus is set by the distance from the nucleus at
which dust reaches its sublimation temperature. This ides is supported by
the observed time delays between variations in the optical and
near-infrared continua (e.g., Fairall~9; Clavel, Wamsteker \& Glass
1989). In more luminous objects this distance is further away and, if the
height of the torus remains constant, the critical angle $\theta_{\rm c}$
must increase, thereby leading to a luminosity dependence in the observed
Type~1 AGN fraction. This is known as the `receding torus' model and was
first suggested by Lawrence (1991).

Of course, the height of the torus need not be the same in all objects but,
since the luminosity dependence of its inner radius is the result of basic
physics, it would be even more surprising if its height varied in such a
way as to produce a constant opening angle. A variation in the Type~1 AGN
fraction with luminosity is therefore to be expected, although the size of
any such variation may be small. If it is large, however, then direct
comparisons between the AGN luminosity function produced by models and the
QSO luminosity function derived from observations (such as that undertaken
by Kauffmann \& Haehnelt 2000) are seriously compromised.

A number of authors have investigated possible evidence for the receding
torus model, frequently using radio-selected AGN samples which are unbiased
in terms of spectral type. These studies (e.g., Hill, Goodrich \& DePoy
1996; Simpson, Rawlings \& Lacy 1999; Willott et al.\ 2000; Simpson \&
Rawlings 2000; Grimes, Rawlings \& Willott 2004) have all supported the
receding torus model, while Simpson (1998) showed how it could explain
apparent differences in supposedly isotropic properties between radio
galaxies and radio-loud quasars with the same radio luminosity. Recent deep
hard X-ray surveys have enabled similar analyses to be performed on samples
of radio-quiet AGN, and the same trend is seen (Ueda et al.\ 2003; Hasinger
2004).

The case for the receding torus model appears strong, although most of the
samples studies so far are relatively small. Alternative models for
unification exist and include warped obscuring discs, broken shells of
material, or an evolutionary scenario where the black hole is at first
smothered but powers a wind which is eventually able to blow away material
and allow the QSO to be directly observed. However, even if the receding
torus model is incorrect, the number of Type 2 AGN need to be determined in
order for AGN to fulfill their cosmological potential. In this paper, we
construct separate [O{\sc~iii}] emission line luminosity functions for
broad and narrow-line AGN from the Second Data Release (DR2) of the SDSS
which we use to investigate how the broad-line fraction varies with
luminosity. The format of the paper is as follows. In Section~2 we describe
how we have produced a sample of emission-line galaxies from the SDSS DR2
catalogue, and in Section~3 we describe the steps taken to produce
luminosity functions for both Type~1 and Type~2 AGN. In Section~4 we derive
the Type~1 AGN fraction and discuss the implications of our
findings. Finally in Section~5 we present a summary. Throughout this paper
we adopt a flat cosmology with $\Omega_{\rm m} = 1-\Omega_\Lambda = 0.3$
and $H_0 = 70\rm\,km\,s^{-1}\,Mpc^{-1}$.

\section{Sample selection}

The SDSS DR2 spectroscopic catalogue covers 2627\,deg$^2$ down to a
Petrosian magnitude of $r'<17.77$ (in addition to colour-selected QSO
candidates which we exclude by means of this magnitude cut).  Spectra are
taken through 3-arcsecond diameter fibres, and so we must impose distance
(i.e., redshift) limits on our sample. The fibres will not encompass the
entire emission line region for nearby AGN, since this can be
$\sim1\rm\,kpc$ in size (e.g., Wilson \& Tsvetanov 1994), while the
dilution by starlight in distant galaxies will make the detection of
emission lines more difficult. Kauffmann et al.\ (2004) show that a
Seyfert~2 sample with a luminosity limit $L_{\rm[OIII]}>10^7\,L_\odot$ and
redshift limits $0.02<z<0.30$ does not suffer from aperture biases, and we
therefore adopt the same constraints for our sample.  This produces a
catalogue of 4304 objects.

We do not apply any reddening corrections to the line luminosities.
This is partly for practical reasons, since there is no reliable way
to determine the reddening to the narrow-line region in broad-line
objects. This is always a problem for studies which aim to compare the
emission line properties of Type~1 and Type~2 AGN, although the fact
that [O{\sc~iii}] emission is seen to be isotropic when no reddening
correction is applied (Mulchaey et al.\ 1994) suggests that
differences between the mean reddening to the two types of galaxy must
be small. Dahari \& De Robertis (1988) estimate that the median
difference is $E(B-V) \approx 0.2$, corresponding to a factor of
$\sim1.7$ at 5007\,\AA. We will address whether this could have an
impact on our results later.

We also do not correct our emission line fluxes for underlying stellar
absorption. Since all our galaxies have very luminous [O{\sc~iii}]
emission lines, they will also have luminous H$\beta$ emission which
is expected to dominate over any absorption. Wake et al.\ (2004) find
that the observed AGN fraction is not sensitive to whether this
correction is made.

\section{Analysis}

\subsection{Spectral classification}

The SDSS spectroscopic pipeline produces a classification for all
objects, based on principal component analysis (Schlegel et al., in
preparation). For our sample of objects, this classification is either
``QSO'' (890 objects) or ``galaxy'' (3414 objects), which might be
expected to correspond to AGN Types 1 and 2, respectively.  However,
while all objects classified as QSOs are expected to display broad
emission lines, the galaxy classification will include starburst
galaxies as well as objects classified as Type~1.8 or 1.9 (where broad
H$\alpha$ emission is seen, but not broad H$\beta$) and possibly some
low-luminosity Seyfert~1s. Since broad emission components will
preclude the use of a line-ratio diagram to discriminate between
starburst and AGN galaxies (Baldwin, Phillips \& Terlevich 1981), we
elect to classify all 4304 galaxies by eye.

It is essential that the classification we choose is not dependent on
the emission line luminosity. Since the broad and narrow emission line
luminosities are strongly correlated, this requires that reddened
objects where the broad emission component is near the limit of
detection be included with the narrow-line objects. Objects which
display broad \textit{wings\/} on the H$\alpha$ line (`Type 1.x'
galaxies) are therefore grouped with the Type~2 galaxies, while those
which have clear broad H$\alpha$ and H$\beta$ emission
\textit{lines\/} are classified as Type~1s.

One object was discovered to have a large noise spike at the expected
location of [O{\sc~iii}]~$\lambda$5007 and was removed from the
sample, since the low luminosities of the visible emission lines and
the absence of the $\lambda$4959 line indicate that its true line
luminosity would be far below our lower limit.

We find that 63 objects classified as QSOs (7.1\,per cent) are not
QSOs at all, but display only narrow emission lines and appear to be
Type~2 galaxies. We therefore reclassify them as such. We also find that
63 objects classified as galaxies (1.8\,per cent) are genuinely QSOs,
with clear broad line emission from both H$\alpha$ and H$\beta$, and
we reclassify these objects. A further 102 (3.0\,per cent) show quite
pronounced broad wings on the H$\alpha$ line only, with a similar
number showing weaker broad wings whose detectability would be a
function of the signal-to-noise ratio in the spectrum. A subsample of
100 objects was classified by Dr C.~Done, and it was only for objects
in this last category where disagreement was found. The total fraction
of SDSS-classified `galaxies' which we believe display broad line
emission is therefore consistent with the figure of 8\,per cent quoted
by Kauffmann et al.\ (2004).

While the reason for the misclassification of Type~2 galaxies as QSOs
is unclear, the Type~1s which are classified as galaxies typically
have either weak broad line emission, or have very narrow permitted
lines ($\sim500\rm\,km\,s^{-1}$). Our interpretation of these as
\textit{bona fide\/} broad-line objects is based on the presence of
Fe{\sc~ii} pseudo-continuum emission between H$\gamma$ and H$\beta$,
the lack of a strong 4000-\AA\ break or Balmer jump, and a continually
rising spectrum at short wavelengths.

\subsection{Separation of Seyfert and starburst galaxies}

We wish to construct comparable Type~1 and Type~2 samples, so that
their luminosity functions (LFs) are properly representative of the
population as a whole. Unfortunately, there is no straightforward way
to do this. The objects we have classified as Type~1s all contain an
AGN component, since the broad emission lines we observe cannot be
produced by stellar processes, but they may also possess a significant
contribution to their emission line luminosities from star formation
activity. Objects where the line emission is almost entirely powered
by star formation may not be included in this sample, however, even if
we have a direct view of the broad line region, since the broad line
emission will be weak. We therefore need to remove pure starburst
galaxies from the sample of narrow-line objects, but also objects
where the AGN contribution is so low that the objects would not have
been classified as Type~1s \textit{if we had been able to see the
BLR.\/}

\begin{figure}
\includegraphics[width=\colwidth]{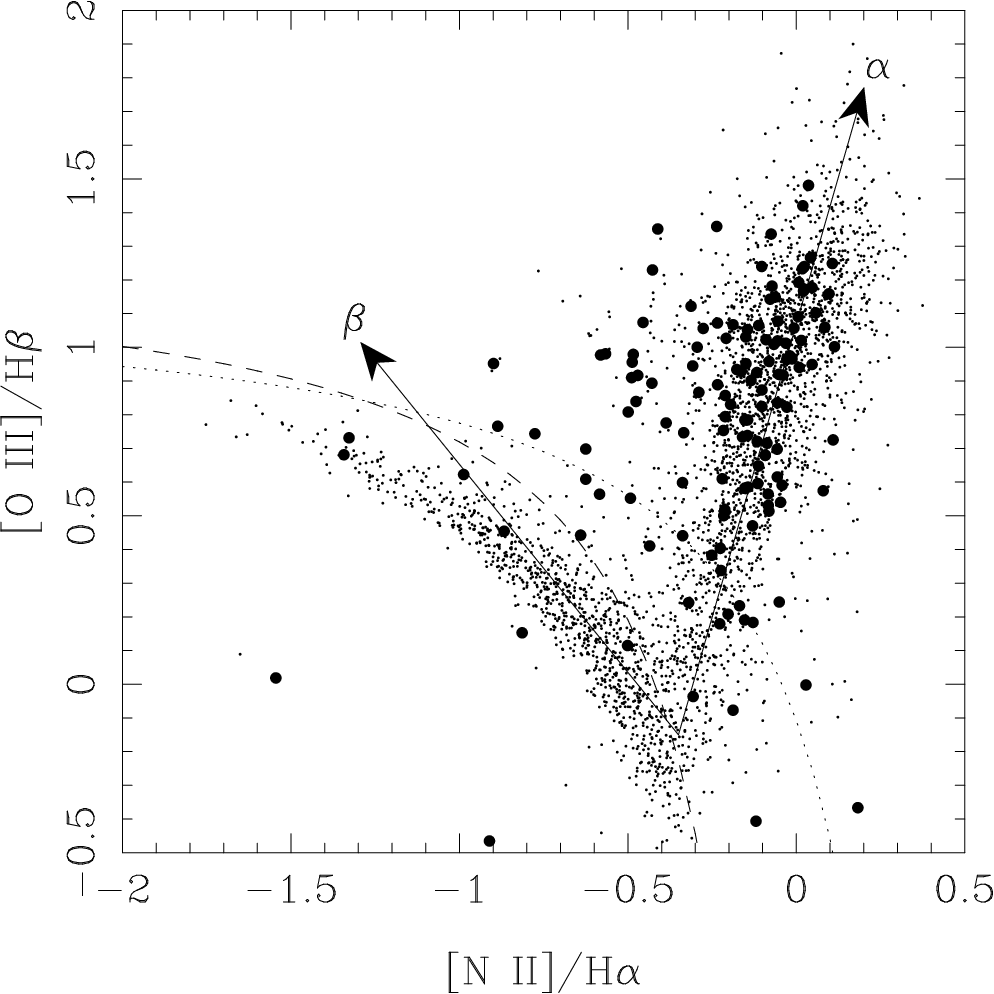}
\caption[]{Emission-line ratio diagram for narrow-line
objects (dots) and Type~1.x galaxies (filled circles). The $\alpha$
and $\beta$ axes represent the ridge lines from the intersection of
the AGN and starburst loci. The dotted and dashed lines are the
classification criteria suggested by Kewley et al.\ (2001) and
Kauffmann et al.\ (2004), respectively.\label{fig:bpt2}}
\end{figure}

Kauffmann et al.\ (2004) use an emission-line ratio diagram of the type
pioneered by Baldwin et al.\ (1981) to separate Seyferts from starburst
galaxies, and we propose a similar method. Since the appropriate diagram
makes use of the narrow Balmer line fluxes, it cannot be used for Type~1
AGN, where these data are unavailable. By plotting the Type~1.x galaxies on
the same diagram, however, we can estimate what regions the Type~1
population would occupy. Fig.~\ref{fig:bpt2} displays these data.

Since our `Type~1.x' classification requires that no broad H$\beta$
emission is seen, the location of these points along the ordinate
should be representative of the Type~1 population. Many of the
Type~1.x galaxies lie in the same region as the Type~2s, because the
broad H$\alpha$ emission is weak and the SDSS measurement therefore
only includes the narrow-line flux. However, the tendency for these
some of objects to have lower [N{\sc~ii}]/H$\alpha$ ratios than the
Type~2 population is clear, as the H$\alpha$ flux measurement for these
includes the broad line emission.  What is also clear is that some
galaxies containing an AGN have low [O{\sc~iii}]/H$\beta$ ratios, and
therefore any cut which excludes starburst galaxies will also exclude
some \textit{bona fide\/} AGN.

We attempt to find the appropriate criteria for separating Type~2
galaxies from pure starbursts by defining two (non-orthogonal) axes as
shown in Fig.~\ref{fig:bpt2}. The axes meet where the loci of the
starburst and AGN sequences meet. The `$\alpha$' axis follows the
ridge line of the AGN sequence, while the `$\beta$' axis follows the
ridge line of the starburst sequence. In this system, the demarcation
criteria proposed by Kewley et al.\ (2001) and Kauffmann et al.\
(2004) correspond to $\alpha\approx0.5$ and $\alpha\approx0$,
respectively.

\begin{figure}
\includegraphics[angle=-90,width=\colwidth]{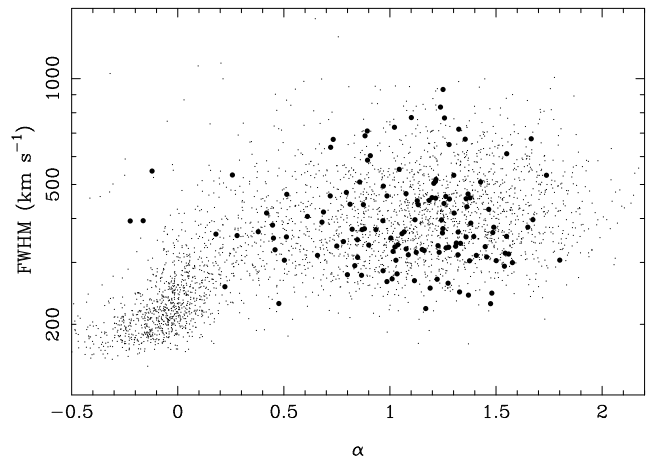}
\caption[]{FWHM of the [O{\sc~iii}] emission lines plotted against
distance along the $\alpha$ axis of
Fig.~\ref{fig:bpt2} for narrow-line objects (dots) and Type~1.x
galaxies (filled circles).\label{fig:fwhm}}
\end{figure}

We also consider the widths of the narrow emission lines (specifically,
[O{\sc~iii}]~$\lambda$5007) to help the differentiation between Seyfert and
starburst galaxies, since it is known that the narrow lines are broader in
Seyfert galaxies than in starbursts (Shuder \& Osterbrock 1981). This is
demonstrated to be the case in Fig.~\ref{fig:fwhm}, where we have plotted
the FWHM of the [O{\sc~iii}] line against $\alpha$ as determined from
Fig.~\ref{fig:bpt2}. For the narrow-line galaxies, we have taken the value
of $\alpha$ from their location in the ($\alpha$,$\beta$) coordinate
system. Since the [N{\sc~ii}]/H$\alpha$ ratio for the Type~1.x galaxies is
affected by broad line emission, we compute values of $\alpha$ for these
galaxies by determining where a galaxy with the same [O{\sc~iii}]/H$\beta$
ratio would lie on the $\alpha$-axis.

The overall distributions of the Type~2 and Type~1.x galaxies are
inconsistent at the $\gg 99.99$\% level, using the two-dimensional
Kolmogorov--Smirnov test of Peacock (1983). However, Fig.~\ref{fig:fwhm}
clearly shows two populations of narrow-line galaxies -- a population with
broad (${\rm FWHM} \ga 250\rm\,km\,s^{-1}$) emission lines extending to
large values of $\alpha$ which appears to be well traced by the Type~1.x
galaxies; and a population of galaxies with narrower emission lines and
$\alpha\approx0$. These two populations can be associated with Seyfert and
starburst galaxies, respectively, and it is necessary to find some way to
discriminate between the two classes.

Since the two distributions obviously differ greatly in the number of
galaxies in the lower left corner of Fig.~\ref{fig:fwhm}, we investigate
discrimination criteria which remove these objects from the narrow-line
sample. We construct a family of four-parameter selection criteria which
require that $\alpha$ and the emission line FWHM are both greater than
certain values, and also that the points satisfy the relation $\log {\rm
FWHM} > m (\alpha_0-\alpha)$ where $m>0$.  We then compare the samples of
narrow-line galaxies selected by each of these criteria with the complete
sample of Type~1.x galaxies in the parameter space of Fig.~\ref{fig:fwhm},
again using a two-dimensional K--S test. We find that the two distributions
can be made consistent at better than 90\,per cent confidence by making a
simple cut with $\alpha>0.20$. No improvement is made by imposing an
additional cut based on line width, and only a very marginal gain results
from adding a further selection criterion which is a function of both
parameters.  This selection lies approximately halfway between the criteria
suggested by Kewley et al.\ (2001) and Kauffmann et al.\ (2004), as
expected. The implication is that the region between their two sets of
criteria represents a mixing line, with our limit being the point at which
the AGN contribution becomes significant (i.e., when it is visible in the
spectrum even if lightly-reddened). In the sample of 4303 emission-line
galaxies there are 851 Type~1 AGN, 141 Type~1.x AGN, 2184 Type~2 AGN, and
1127 starburst galaxies (using the $\alpha$ criterion to classify the
latter two classes).

\subsection{Comparison of luminosity functions}

Since the SDSS DR2 spectroscopic survey is a complete magnitude-limited
survey, Type~1 galaxies will appear in relatively greater numbers because
of the additional flux from the nuclear non-stellar continuum. We can
however account for this by constructing LFs using the $1/V_{\rm a}$
estimator (Schmidt 1968; Rowan-Robinson 1968).  We then compare the Type~1
LF with the LFs of three subsamples of Type~2 galaxies, constructed by
applying different selection criteria to the entire sample in an attempt to
exclude the starburst galaxies. Two of these subsamples, named Type~2-Ka
and Type~2-Ke, are produced by applying the line ratio selection criteria
of Kauffmann et al.\ (2004) and Kewley et al.\ (2001), respectively, while
the third, named Type~2-$\alpha$, is constructed using the $\alpha>0.2$
criterion derived above.

\begin{figure}
\includegraphics[angle=-90,width=\colwidth]{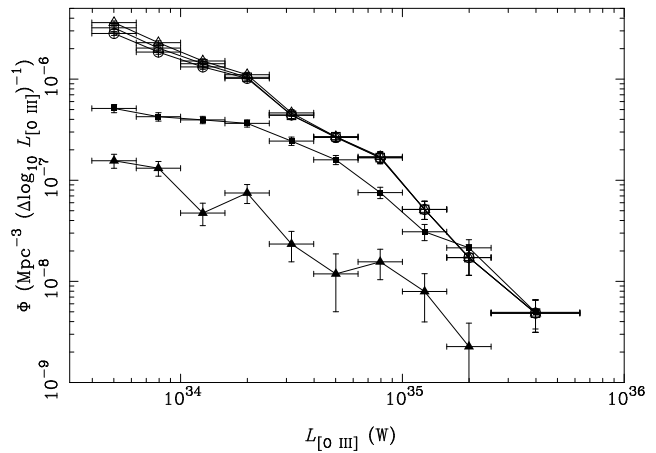}
\caption[]{Luminosity functions for Type~1 galaxies (filled squares) and
samples of Type~2 galaxies selected from the narrow-line objects using the
criteria of Section~3.2 (open squares), Kewley et al.\ (2001; open
circles), and Kauffmann et al.\ (2004; open triangles). Also shown
(filled triangles) is the luminosity function of the Type~1.x galaxies,
although these are also included in all the Type~2 samples.\label{fig:lf}}
\end{figure}

The three Type~2 LFs are compared with the Type~1 LF in
Fig.~\ref{fig:lf}. It is clear that, although the numbers of each type of
Seyfert are similar at $L_{\rm[O\,III]} \ga 10^{34.5}\rm\,W$, Type~2s
outnumber the Type~1s at lower luminosities by a factor of
$\sim5$. Although there are differences between the numbers of lower
luminosity Type~2s in the three samples, these differences are small
compared to the difference between these samples and the number of Type~1s.

The different shapes of the LFs cannot be explained
by a difference in the mean reddening of the Type~1 and Type~2 samples,
since a reddening correction is simply a horizontal shift. The
different shapes also cannot be due to problems with the
Seyfert/starburst separation. The criterion defined by Kewley et al.\
(2001) is conservative in the sense that it should exclude
\textit{all\/} starburst galaxies, so the excess of Type~2 galaxies over
Type~1 galaxies at lower [O{\sc~iii}] luminosities must be genuine,
since even the Type~2-Ke sample exceeds the Type~1 sample. Conversely, the
difference also cannot be due to an artificial deficit of high
luminosity Type~2 galaxies, since all three samples have effectively the
same LFs at $L_{\rm[O\,III]} \ga 10^{34.5}$\,W.

A final possibility is perhaps that a significant fraction of the
lower luminosity Type~1 galaxies have been misclassified, presumably
because the broad lines are weak and below the detection
threshold. This seems unlikely since the broad lines are much stronger
than the narrow lines. In our Type~1 sample, the median
[O{\sc~iii}]/H$\alpha$ ratio is 0.15, while 90\,per cent of objects
have a ratio less than 0.5. We therefore expect all but a small
fraction of Type~1 galaxies to have readily identifiable broad H$\alpha$
emission (although the [O{\sc~iii}] lines in our sample only need to
be detected at ${\rm S/N} > 5$, most are detected at much higher
significance -- all but 20 are detected at ${\rm S/N} > 10$, and over
half have ${\rm S/N} > 20$). Furthermore, we also display the LF of
the Type~1.x galaxies in Fig.~\ref{fig:lf}, and there is clearly no bias
against lower luminosity objects. We therefore conclude that the
luminosity functions of Type~1 and Type~2 galaxies are genuinely
different.

For the remainder of this paper, we use only one Type~2 luminosity
function. We adopt the Type~2-$\alpha$ sample, and add in quadrature to
the Poisson uncertainty the standard deviation of the three samples.

\section{Discussion}

\subsection{The Type 1 fraction}

The different luminosity functions for our samples of Type~1 and Type~2
galaxies can also be described as a luminosity dependence of the
Type~1 AGN fraction. In the context of the receding torus model, such
a change is described by
\begin{equation}
f_1 = 1 - [1+3L/L_0]^{-0.5}
\label{eq:RT}
\end{equation}
(e.g., Simpson 1998), where $L_0$ is the luminosity at which the
numbers of Type~1 and Type~2 AGN are equal (i.e., an opening angle of
$\theta_0 = 60^\circ$). The `luminosity' of concern here is the
ultraviolet--optical radiation which heats the dust, which is
difficult to measure directly, but the constant equivalent width of
the [O{\sc~iii}] emission line in QSOs (e.g.\ Miller et al.\ 1992; see
also Simpson 1998) suggests that it should be a good tracer. Since the
receding torus model is able to explain a wide range of observed AGN
properties (Simpson 2003 and references therein; see also Grimes et
al.\ 2004), we investigate whether it is also able to explain the
results of the previous section.

\begin{figure}
\includegraphics[angle=-90,width=\colwidth]{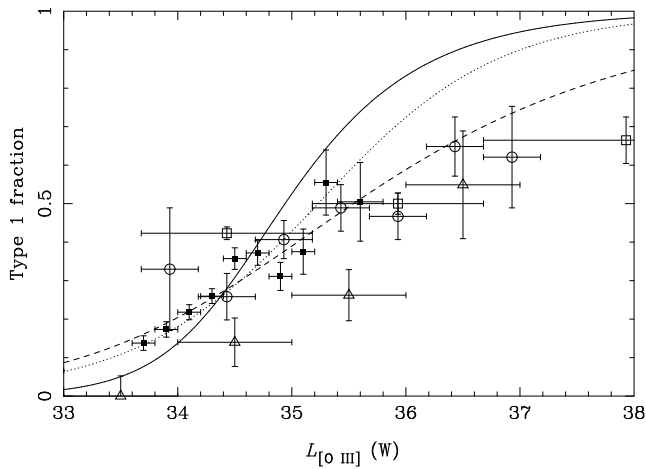}
\caption[]{Type~1 AGN fraction as a function of [O{\sc~iii}]
luminosity, from this paper (solid squares), Ueda et al.\ (2003; open
squares), Hasinger (2004; open circles), and Grimes et al.\ (2004;
open triangles). The data from Ueda et al.\ and Hasinger have been
converted to $L_{\rm[O~III]}$ as described in the text. The solid line
shows the best-fitting `standard' receding torus model
(Equation~\ref{eq:RT}) while the dashed line shows the best-fitting
model where the torus height is allowed to vary
(Equation~\ref{eq:RTh}), and the dotted line shows the best-fitting
model where the proportionality between $L_{\rm[O~III]}$ and $L_{\rm
rad}$ is broken (Equation~\ref{eq:RTf}).\label{fig:qsofrac}}
\end{figure}

We complement our data from similar analyses in the literature, which
typically extend to higher luminosities. Both Ueda et al.\ (2003) and
Hasinger (2004) measure the Type~1 AGN fraction in hard X-ray selected
samples, which should be relatively free from orientation biases except for
Compton-thick AGN, whose existence will case the Type~1 fraction to be
systematically overestimated. About half of nearby low luminosity Type~2
AGN are Compton-thick (Risaliti et al.\ 1999) and, if this fraction is true
at all luminosities then the Type~1 fraction will be overestimated by about
50\,per cent. We do not make any correction for Compton-thick objects but
note that the Type~1 fractions derived from X-ray-selected samples should
be considered as upper limits to the true fractions. We convert the
2--10\,keV luminosity (corrected for intrinsic absorption) used by these
authors into [O{\sc~iii}] luminosity using the mean ratio for Seyfert
galaxies from Mulchaey et al.\ (1994), i.e.,
\begin{equation}
L_{\rm[O~III]} = 0.015 \times L_{\rm2-10keV} \, .
\end{equation}
We also add the data of Grimes et al.\ (2004) from the 3CRR, 6CE, and 7C
flux-limited samples of radio sources. The results are shown in
Fig.~\ref{fig:qsofrac}, where a clear and highly significant ($>99.99$\,per
cent confidence) correlation between the Type~1 fraction and luminosity is
seen for which the receding torus model provides an attractive, but not
unique, explanation for this correlation. We note that the points from
Grimes et al.\ (2004) tend to be slightly lower than those from the other
works and this might reflect a difference in the torus properties of
radio-loud and radio-quiet AGN, although the discrepancy is not highly
significant.

Although the four samples plotted in Fig.~\ref{fig:qsofrac} all show
the same trend, we concentrate for now on the data from our analysis
alone since there are less well understood selection effects in the
X-ray samples (see Treister et al.\ 2004). These cover more than two
orders of magnitude in $L_{\rm[O~III]}$ but are not well-fit by the
standard receding torus model, with the best fit having $\chi^2/\nu =
65.8/9$ for $L_0 = 10^{34.90}\rm\,W$.

We therefore suggest a modification to the standard receding torus
model, where the height of the torus depends on the luminosity of the
AGN. Such a dependence could plausibly be modulated via the mass of
the black hole and the strength of its associated gravitational
potential. We parametrize this dependence as $h \propto L^\xi$, and
Equation~\ref{eq:RT} therefore becomes
\begin{equation}
f_1 = 1 - [1+3(L/L_0)^{1-2\xi}]^{-0.5} \, .
\label{eq:RTh}
\end{equation}

This produces a more acceptable fit ($\chi^2/\nu = 12.1/8$) for $L_0
= 10^{35.37}$\,W and $\xi=0.23$, which we also show in
Fig.~\ref{fig:qsofrac}. Arshakian (2004) made a similar
parametrization when studying the luminosity-dependent variations of
QSO fraction and projected linear size in a radio source sample. He
defined a parameter $A$ such that $\tan \theta_{\rm c} \propto L^A$
(therefore $A \equiv 0.5-\xi$) and found $A_{\rm s} = 0.35\pm0.09$ and
$A_{\rm f} = 0.26\pm0.02$ when considering the linear sizes and QSO
fractions, respectively. The latter value is in excellent agreement
with the best-fitting value of $\xi$ we determine here.

An alternative explanation for this modification is that
$L_{\rm[O\,III]}$ is not a good substitute for $L_{\rm rad}$, the
radiative luminosity which heats the dust in the torus. If
$L_{\rm[O~III]} \propto L_{\rm rad}^{1-2\xi}$, then the standard
receding torus model would fit the data. However, such an extreme
disproportionality can be excluded based on a number of arguments such
as the near-constancy of [O{\sc~iii}] equivalent width in QSOs,
(Miller et al.\ 1992), the proportionalities between [O{\sc~iii}]
luminosity and other quantities (e.g., Rawlings \& Saunders 1991;
Mulchaey et al.\ 1994; Grimes et al.\ 2004), and the dependence of
[O{\sc~iii}] emission on ionization parameter (Simpson 1998).

A more subtle variation in the relationship between $L_{\rm rad}$ and
$L_{\rm[O~III]}$ may provide an explanation. In the standard unified model
for AGN, only a fraction $\sim f_1$ of the ionizing radiation can escape to
illuminate the narrow-line region on scales $>1$\,pc. This explains the
(bi)conical emission-line structures seen in some Seyfert galaxies (e.g.\
Pogge 1989; Haniff, Wilson \& Ward 1991; Simpson et al.\ 1997), and
suggests that $L_{\rm[O~III]} \propto L_{\rm rad} f_1$. In such a scenario,
\begin{equation}
f_1 = 1 - [1 + 1.5 L_{\rm[O~III]}/(L_0 f_1)]^{-0.5} \, ,
\label{eq:RTf}
\end{equation}
which can be rearranged to give a cubic equation in $f_1$ whose roots
are functions of $L_{\rm[O~III]}$. This model is also shown in
Fig.~\ref{fig:qsofrac} and provides an acceptable fit (i.e.\ cannot be
rejected at 95\,per cent confidence) to our data ($\chi^2/\nu = 16.0/9$
for $L_0=10^{35.19}$\,W).

Fig.~\ref{fig:qsofrac} suggests that we should favour the torus described
by Eqn~\ref{eq:RTh} over that from Eqn~\ref{eq:RTf} by virtue of the
quality of fit to the X-ray datapoints. However, we caution against leaping
to such a conclusion and suggest that some of the most luminous QSOs may
have been misclassified as Type~2 objects. While this seems
counterintuitive, such objects will inevitably be at the highest redshifts
and therefore the optical spectroscopy performed on them will sample the
rest-frame ultraviolet. Even relatively small amounts of dust could be
sufficient to extinguish the broad emission lines and give such objects the
appearance of a Type~2 AGN in the UV. While there should be very few
luminous AGN which are lightly reddened because the observer's line of
sight just grazes the torus (see Hill et al.\ 1996), the required
extinctions may arise from dust in the host galaxy which is not associated
with the active nucleus. The repeated reclassifications of `Type~2 QSOs'
after near-infrared spectroscopy reveals broad H$\alpha$ emission (e.g.,
Halpern, Eracleous \& Forster 1998; Halpern, Turner \& George 1999;
Akiyama, Ueda \& Ohta 2002) suggests that there may be some truth in
this. On the other hand, the QSO fraction may be overestimated if there
exists a significant population of luminous AGN with Compton-thick tori
which are not detected in hard X-ray surveys. What is required is a
systematic near-infrared study of either spectroscopy of imaging, similar
to that which Simpson et al.\ (1999) performed for $z\sim1$ powerful radio
sources. This should be possible in the very near future with the
combination of
\textit{Chandra\/} or \textit{XMM--Newton\/} and \textit{Spitzer\/}
imaging in a number of fields.

Having suggested that both modifications to the receding torus model
fit the data equally well, we return to the issue of whether breaking
the proportionality between the ionizing and [O{\sc~iii}] emission-line
luminosities is in conflict with observations. First, this does not
affect the isotropy of the [O{\sc~iii}] emission, since the
emission-line luminosities of Type~1 and Type~2 AGN with similar
ionizing luminosities should still be the same. However, it predicts
that the observed equivalent width of [O{\sc~iii}] should be
proportional to $f_1$ (i.e., increase with increasing AGN luminosity),
whereas observations show it to be constant with luminosity (Miller et
al.\ 1992). However, over the range of luminosities studied by Miller
et al.\ (approximately two orders of magnitude), $f_1$ is predicted to
vary by a factor of about 2, which is equal to the $1\sigma$ scatter
in the rest-frame equivalent width. Furthermore, this effect will be
lessened by those objects seen at small polar angles where the
continuum, but not the emission-line, luminosity is enhanced. A
quantitative analysis of this effect depends on the assumed angular
and wavelength dependence of the ionizing continuum and is therefore
rather model-dependent and beyond the scope of this paper. However, we
do predict that the scatter in the equivalent width of the
[O{\sc~iii}] line should increase with luminosity.

\subsection{Implications for AGN evolution}

Studies of AGN evolution have so far primarily relied on samples of QSOs
since these objects are readily visible out to large distances. Obviously
such samples contain only a fraction of the AGN at any redshift or
luminosity since Type~2 objects do not meet the colour and/or morphological
selection criteria (while hard X-ray surveys are less biased against Type~2
AGN, they presently lack the combination of depth and area coverage to
properly sample the AGN luminosity function in an unbiased manner). We
therefore wish to determine how the luminosity-dependent QSO fraction might
affect the conclusions drawn from studies of QSO evolution.

Since there is a significant stellar component to the broad-band
magnitudes of our Type~1 galaxies which is not the case for observed
samples of QSOs, we estimate the non-stellar continuum luminosities
from our [O{\sc~iii}] luminosities.We assume a mean rest-frame
equivalent width (relative to the non-stellar continuum) for the
$\lambda$5007 line of $\epsilon=24$\,\AA\ (Miller et al.\ 1992), and a
spectral shape $S_\nu \propto \nu^{-0.44}$ (Vanden Berk et al.\ 2001;
the result is extremely insensitive to the shape of the spectrum due
to the small wavelength difference between the [O{\sc~iii}] line and
the $B$ band) in the rest-frame optical. It then follows that
\begin{equation}
M_B = -22.0 - 2.5 \log (L_{\rm [O\,III]}/10^{35}\,{\rm W})
+ 2.5 \log (\epsilon/24{\rm \,\AA})
\label{eq:mb}
\end{equation}
and $M_{b_J}=M_B-0.09$ for the canonical QSO spectrum (Blair \&
Gilmore 1982).

Our Type~1 LF can be well-fit with the two power-law parametrization of the
2QZ team,
\begin{equation}
\Phi(L) = \frac{2\Phi(L^*)}{(L/L^*)^{-\alpha}+(L/L^*)^{-\beta}}
\end{equation}
(e.g., Croom et al.\ 2004; note that the normalization factor of 2 is
missing from papers by this team). A good fit ($\chi^2/\nu = 4.8/6$)
is found with the parameters $\alpha=-2.86$, $\beta=-1.06$, $L^* = 3.5
\times 10^{34}\rm\,W$, and $\Phi^* = 2.24 \times
10^{-7}\rm\,Mpc^{-3}\,dex^{-1}$. The power law indices are similar to
those derived by Croom et al.\ (2004), who found $\alpha=-3.25$ and
$\beta=-1.01$. However, a better comparison would be to the combined
(i.e., Types 1 and 2) Seyfert LF since, while the Croom et al.\ (2004)
LFs are relatively insensitive to the lack of Type~2 objects, our
Type~1 LF is a poor representation of the overall AGN LF. The fit to
the combined Seyfert luminosity function is improved ($\chi^2/\nu =
3.3/6$), primarily due to the larger error bars resulting from the
uncertain classification of narrow-line objects, with $\alpha=-3.11$,
$\beta=-1.59$, $L^* = 4.4 \times 10^{34}\rm\,W$, and $\Phi^* = 5.25
\times 10^{-7}\rm\,Mpc^{-3}\,dex^{-1}$.

There is moderately good agreement between the values of $M^*$ and
$\Phi^*$ from our combined Seyfert sample and the extrapolation of the
exponential pure luminosity evolution of Croom et al.\ (2004). We
obtain $M_{b_J}^* = -21.20$ and $\Phi^* = 1.31 \times
10^{-6}\rm\,Mpc^{-3}\,mag^{-1}$ compared to $M_{b_J}^* = -21.46$ and
$\Phi^* = 1.84 \times 10^{-6}\rm\,Mpc^{-3}\,mag^{-1}$ from their work
(the agreement is less good for the polynomial evolution, where a
value of $M_{b_J}^*=-22.15$ is predicted for our sample). The lower
space density we derive cannot be due to incompleteness in our sample
of Seyfert galaxies since our spectroscopic selection will include all
such objects (accounting for the uncertainty in the starburst--Seyfert
classification), whereas the optical colour selection used to select
QSOs will exclude a population of red QSOs which may exist in
significant numbers (Webster et al.\ 1995; Benn et al.\ 1998; Richards
et al.\ 2003).

\begin{figure}
\includegraphics[angle=-90,width=\colwidth]{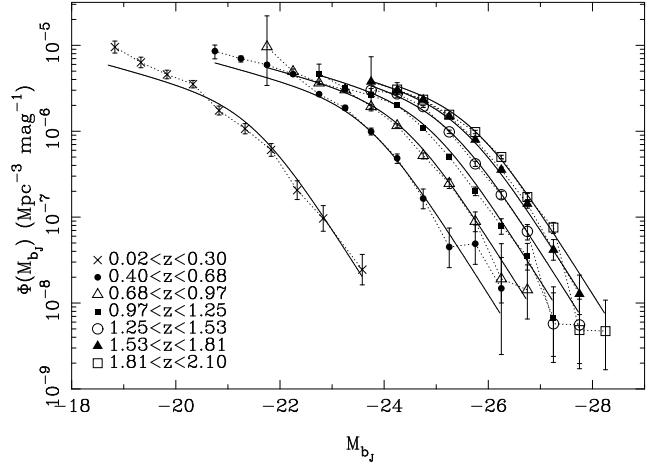}
\caption[]{Our Seyfert (Types 1 and 2) luminosity function converted
to $M_{b_J}$ (crosses) and plotted with the modified QSO optical luminosity
functions of Croom et al.\ (2004; all other symbols). A two power-law model
with pure luminosity evolution has been fitted to the data, as described in
the text, and the results are shown by the solid curves.\label{fig:croom}}
\end{figure}

We also try to simultaneously fit our Seyfert LF and the six QSO LFs
of Croom et al.\ (2004), after using the modified receding torus model
to account for the missing Type~2 objects. We use their two power-law
model with an exponential form for the luminosity evolution, although
we fit the value of $M^*$ for our Seyfert LF as a separate parameter
since an extremely poor fit is obtained otherwise. Following Croom et
al., we use only the $M_{b_J}<-22.5$ bins, and we show the best fit in
Fig.~\ref{fig:croom}. While this fit is unacceptably poor in a formal
sense ($\chi^2/\nu = 175/58$), it does reproduce the gross
characteristics of the evolving LF and, in particular, the power law
indices differ from those derived from the Type~1 (QSO) LF alone.

Unsurprisingly, the value of $\alpha$ is least affected. From our earlier
calculation that $L_0 = 10^{35.53}$\,W, we conclude that QSOs outnumber
Type~2 objects where $M_B \la -23.3$ and therefore the bright end of the
luminosity function is not strongly affected by the absence of Type~2
objects and the power law slope is only moderately steepened by their
inclusion -- we find $\alpha=-3.42$ as opposed to $\alpha=-3.25$. However,
the faint end slope is made much steeper ($\beta=-1.41$ compared to
$\beta=-1.01$), and is closer to the faint-end slope of the hard X-ray
luminosity function ($\beta=-1.86$; Ueda et al.\ 2003). This steepening has
a large effect on the contribution from faint AGN to the overall AGN
luminosity.  Integrating the two power-law model, the total luminosity from
AGN with $L>L^*$ is $\sim0.4\Phi^*L^*$. The total luminosity from AGN
fainter than $L^*$ does not converge for $\beta<-1$ and so is dependent on
the low-luminosity cutoff, but in the decade below $L^*$, it is
$\sim2.3\Phi^*L^*$ for $\beta\approx-1.0$ and nearly twice as much for
$\beta\approx-1.5$. Wolf et al.\ (2003) find a downturn in the shape of the
LF at faint absolute magnitudes ($M_B>-22$, not probed by 2QZ) which allows
the luminosity density integral to converge. However, the increased number
of Type~2 AGN predicted at these luminosities significantly affects the
strength of this downturn and therefore the contribution by faint AGN to
the overall luminosity density. Such objects would be fainter than the
low-luminosity Type~1 AGN studied by Wolf et al.\ (2003) and therefore
could not have yet been spectroscopically identified in deep X-ray surveys
such as Barger et al.\ (2003). Given that a putative further population of
Compton-thick Type~2 AGN would not even appear in such X-ray surveys, there
is a distinct likelihood that the importance of low luminosity AGN has been
significantly underestimated.

Finally, while attempting to tie the present-day black hole mass density to
the evolution of QSOs, Yu \& Tremaine (2002) note that their results are
dependent on the prevalence of obscured low-luminosity AGN. In particular,
the accretion efficiency can be made roughly independent of AGN luminosity
if the obscured AGN fraction is larger for low mass black holes, and hence
less luminous AGN, as predicted by the receding torus model.

\subsection{The Cosmic X-ray Background}

While the predictions of the receding torus model agree well with the
observed number of X-ray-selected Type~2 AGN, models for the CXB (e.g.,
Comastri et al.\ 1995; Wilman, Fabian \& Nulsen 2000) require the presence
of an additional Compton-thick population which is not detectable in such
surveys. As discussed previously, this means the true Type~1 fraction is
lower than the observations suggests, and hence there is an apparent
discrepancy between the receding torus model and models for the CXB.

However, the number of Compton-thick AGN proposed is a relatively small
fraction of the total AGN population. Ueda et al.\ (2003) fit the CXB by
proposing that there are as many AGN with $10^{24}$\,cm$^{-3} < N_{\rm H} <
10^{25}$\,cm$^{-3}$ as with $10^{23}$\,cm$^{-3} < N_{\rm H} <
10^{24}$\,cm$^{-3}$, but there is a strong luminosity dependence in the
shape of the $N_{\rm H}$ distribution. Therefore, while there are many
Compton-thick AGN at low redshift/luminosity (Risaliti et al.\ 1999), there
are many fewer at the redshift/luminosity which produces the bulk of the
CXB.

In addition, Type~1 AGN with significant X-ray obscuration are known to
exist, the original member of this class being MR~2251$-$178 (Halpern 1984;
see also Page et al.\ 2004), so the Type~1 fraction need not be the same as
the fraction of X-ray unabsorbed AGN. Although the column densities of
these objects tend not to exceed $\sim 10^{22.5}\rm\,cm^{-2}$, Willott et
al.\ (2003, 2004) have found objects with greater absorption ($N_{\rm
H}>10^{23}\rm\,cm^{_2}$) whose optical/infrared spectra are consistent with
lightly-reddened QSOs. It is therefore possible that the most heavily
absorbed AGN also suffer significant non-torus reddening and may have
escaped detection in both X-ray and optical QSO surveys.

Finally, a large discrepancy between the receding torus model and the
observations of hard X-ray-selected AGN occurs only at very high
luminosities ($L_{\rm[OIII]}>10^{36}$\,W, or $L_{\rm X}>10^{38}$\,W) and
this population contributes almost negligibly to the overall CXB (e.g.,
figs~16 and 17 of Ueda et al.\ 2003). We conclude therefore that the
receding torus model does not conflict with our current understanding of
the Cosmic X-ray Background.

\section{Summary}

We have constructed separate luminosity functions for Seyfert~1 and
Seyfert~2 galaxies, using the Second Data Release of the Sloan Digital Sky
Survey. We have shown that these luminosity functions have different
shapes, resulting in an increase in the Type~1 AGN fraction with luminosity
which is in quantitative agreement with that determined from hard
X-ray-selected samples. We show that the form of this luminosity dependence
is inconsistent with the standard receding torus model, and propose a
modification where the height of the torus varies with luminosity as $h
\propto L^{0.26}$, although we note that a more robust spectral
classification of distant luminous AGN is desired. Assuming that this
modification to the receding torus model provides a reasonable correction
factor to account for the `missing' Type~2 AGN in QSO surveys, we have
shown that the contribution to the cosmic accretion luminosity from faint
AGN may have been significantly underestimated.

\section*{Acknowledgments}

I would like to thank Matt Jarvis for useful discussions, Chris Done for
agreeing to classify a subsample of spectra, and the referee, Andy
Lawrence, for his helpful report. I also wish to thank PPARC for financial
support in the form of an Advanced Fellowship.

Funding for the creation and distribution of the SDSS Archive has been
provided by the Alfred P. Sloan Foundation, the Participating
Institutions, the National Aeronautics and Space Administration, the
National Science Foundation, the U.S. Department of Energy, the
Japanese Monbukagakusho, and the Max Planck Society. The SDSS Web site
is http://www.sdss.org/.

The SDSS is managed by the Astrophysical Research Consortium (ARC) for
the Participating Institutions. The Participating Institutions are The
University of Chicago, Fermilab, the Institute for Advanced Study, the
Japan Participation Group, The Johns Hopkins University, Los Alamos
National Laboratory, the Max-Planck-Institute for Astronomy (MPIA),
the Max-Planck-Institute for Astrophysics (MPA), New Mexico State
University, University of Pittsburgh, Princeton University, the United
States Naval Observatory, and the University of Washington.

\end{document}